# Adaptive Predictive Portfolio Management Agent

Anton Kolonin[1,2][0000-0003-4180-2870], Alexey Glushchenko[1][0000-0002-8183-9208], Arseniy Fokin[1][0000-0002-7868-6482], Marcello Mari[1], Mario Casiraghi[1], Mukul Vishwas[1][0000-0002-8824-1954]

[1] SINGULARITYDAO LABS DMCC, Dubai, United Arab Emirates
[2] Novosibirsk State University, Novosibirsk, Russian Federation
akolonin@gmail.com

**Abstract.** The paper presents an advanced version of an adaptive market-making agent capable of performing experiential learning, exploiting a "try and fail" approach relying on a swarm of subordinate agents executed in a virtual environment to determine optimal strategies. The problem is treated as a "Narrow AGI" problem with the scope of goals and environments bound to financial markets, specifically crypto-markets. Such an agent is called an "adaptive multi-strategy agent" as it executes multiple strategies virtually and selects only a few for real execution. The presented version of the agent is extended to solve portfolio optimization and re-balancing across multiple assets so the problem of active portfolio management is being addressed. Also, an attempt is made to apply an experiential learning approach executed in the virtual environment of multi-agent simulation and backtesting based on historical market data, so the agent can learn mappings between specific market conditions and optimal strategies corresponding to these conditions. Additionally, the agent is equipped with the capacity to predict price movements based on social media data, which increases its financial performance.

## 1    Introduction

The approach and architecture of an adaptive agent acting in an environment of the financial market, being a "Narrow Artificial General Intelligence" (Narrow AGI) agent specialized in the financial domain, has been actively discussed in recent years [Raheman, 2023]. It was initially proposed as an agent-based solution for active portfolio management, and the overall architecture was outlined [Raheman, KNOTH 2021]. The latest work has explored the possibility of an AGI agent learning the ability for financial market prediction [Oswald, 2023].

Some earlier works, such as [Tsantekidis, 2018] and [Ganesh, 2019], have approached the use of machine learning for the specific problem of market-making based on the limit order book on centralized exchanges in conventional financial mar-kets. Other later works, such as [Sadighian, 2019] and [Sadighian, 2020], have tried to narrow this down by using reinforcement learning applied to the crypto-market.

The idea of the so-called "adaptive multi-strategy agent" (AMSA) was introduced in [Raheman, AGI 2021]. In this approach, the market-making agent performs purposeful activity [Vityaev, 2015] targeting the maximization of financial returns by means of experiential learning [Kolonin, 2021] through a "try and fail" approach. It relies on a swarm of subordinate agents being executed in a virtual environment to determine optimal strategies, which are then executed in the real environment, as shown in Fig. 1. Such an agent is called an "adaptive multi-strategy agent" as it executes multiple strategies virtually and selects only a few for real execution. The virtual environment for strategy evolution is created with multi-agent simulation of the real market based on either a) a completely synthetic population of agents playing roles of market-makers and traders driven by the historical price curve or b) backtesting by simulation of exchange operation matching historical records of real trades executed on the market against historical snapshots of the limit order book (LOB) structure. The latest developments of this approach were presented recently [Raheman, 2023], showing the capacity of this approach to perform in volatile crypto-markets.

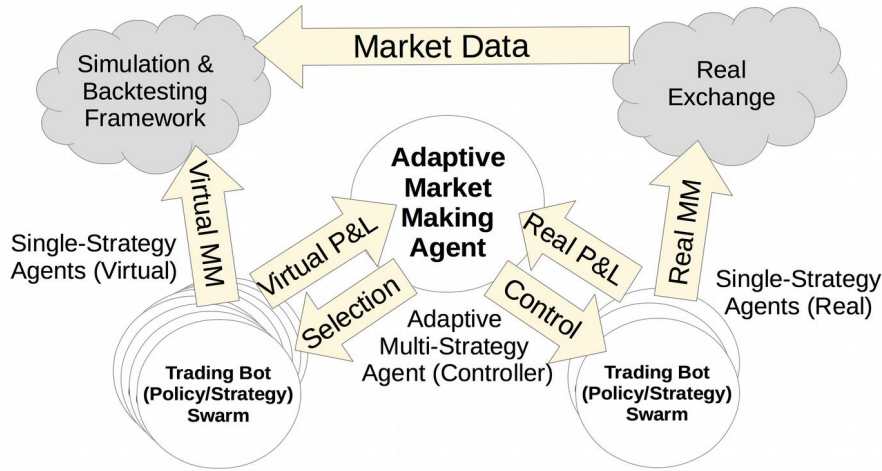

**Fig. 1.** Architecture of the "adaptive multi-strategy agent" for market-making (MM). Market data, including records of executed trades and snapshots of the limit order book structure, are collected by a simulation and backtesting framework (at the top). The "controller" agent runs a swarm of trading bots that execute a wide range of market-making strategies in a virtual environment, returning virtual profits and losses (P&L) associated with these strategies (on the left). On every strategy evaluation cycle, the "controller" selects the top-performing (in terms of P&L) strategies for a given market- momentum and creates another smaller swarm of market-making bots to execute the selected strategies on a real exchange to collect real P&L (on the right).

The environment of the AMSA agent consists of market data [Raheman, AGI 2021] as well as social media data [Kolonin, 2023], which can also be used for price movement prediction. The study of sentiment analysis for the purpose of market price prediction has been explored before in [Deveikyte, 2020] and [Vishwas, 2022], but the latest study [Kolonin, 2023] suggests "cognitive distortions," known in cognitive psychology, may serve as indicators of manipulations and panic.

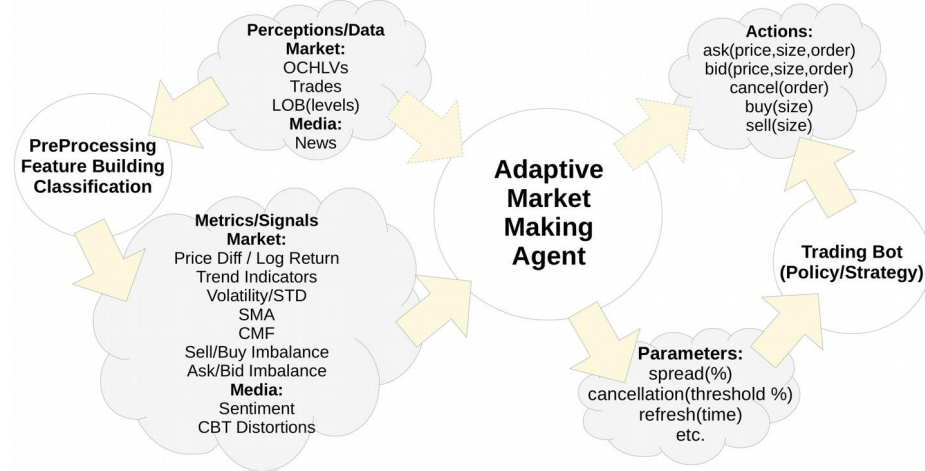

**Fig. 2.** Operational space of the adaptive market-making agent as a "Narrow AGI" operating in an environment represented by financial market data, relevant social media news feeds, and performing financial transactions on the market according to strategies defined by specific parameters.

## 2 Advanced agent architecture

Architecture of the AMSA agent explored in this study extends the one suggested in earlier work [Raheman, 2023], as shown in Fig. 2. The agent presented in this study is capable of perceiving not only market data but also social media data. In order to optimize performance, the data is not consumed directly but is pre-processed. The raw market data, such as open-high-low-volume frames, raw trades, and LOB snapshots, are converted into about two hundred derivative metrics as time series, including derivatives and imbalances between buy and sell volumes or between volumes of buy/ sell trades and ask/bid limit orders. In turn, the social media data is processed so that social media and cognitive distortion metrics are identified and turned into time series as well, according to [Vishwas, 2022] and [Kolonin, 2023].

The parameters of an agent strategy used in this work were slightly different compared to the ones used in earlier works [Raheman, AGI 2021] and [Raheman, 2023]. We still use the percentage of the spread between the bid and ask prices of the limit orders along with the order refresh rate. But we have replaced the "order cancellation policy" (with only three fixed policies used) used in the above-mentioned studies with a "cancellation threshold" that specifies what the magnitude of the price movement should be in order to have the orders re-created. The latter provides more granularity and accuracy for strategy identification.

In addition to the extended version of the AMSA agent, an attempt was made to apply the experiential learning approach [Kolonin, 2021] executed in the virtual environment of multi-agent simulation and backtesting based on historical market data so that the agent could learn mappings between specific market conditions and optimal strategies corresponding to these conditions.

Moreover, we explored how the entire principle of the adaptive multi-strategy operations can be adopted for a generic case of active portfolio management, including portfolio optimization and rebalancing across multiple assets, as illustrated by Fig. 3. For this purpose, we extended the agent design in two ways. First, we made it possible to evaluate, by means of simulation and backtesting, all "candidate" strategies across different markets, so the allocation of portfolio funds can be seen in a two-dimensional space with assets or instruments on one axis and a specific strategy, identified by its parameters, on the other axis. It should be noted that in our experiments described below, all assets/instruments were traded against the USDT currency.

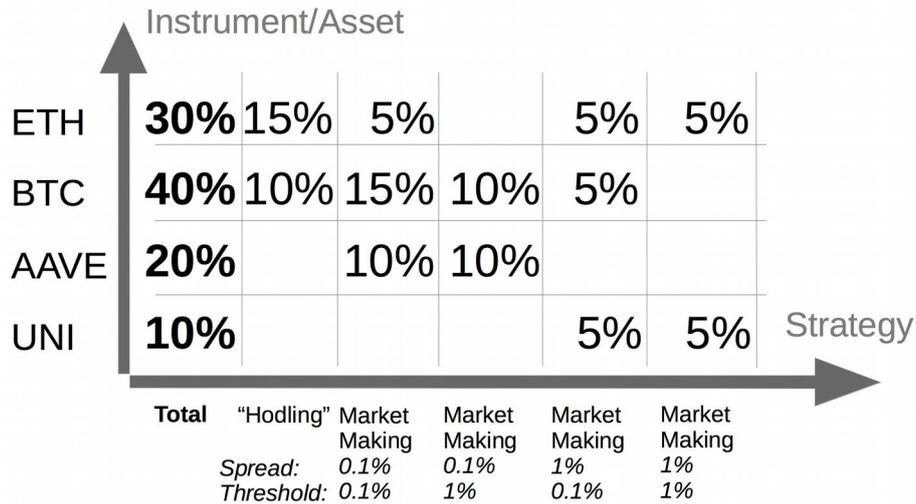

**Fig. 3.** Two-dimensional space for fund allocation in adaptive multi-asset and multi-strategy portfolio management. An asset in this case is a cryptocurrency, and a strategy can either be "hodling," which involves locking funds in an asset for the period of strategy evaluation or execution, or market-making with specific values such as bid/ask spread or order cancellation threshold.

In our experiment design, we extended the funds allocation to be unevenly distributed across both the assets and the strategies within a single asset. This allowed the amount of funds on a strategy execution cycle to be proportional to the positive returns observed on the previous strategy evaluation cycle, which was found to be beneficial.

In summary, the agent architecture we explored can be called adaptive predictive active portfolio management based on multiple strategies, being concurrently executed in the virtual environment of simulation and backtesting. The selected strategies are subsequently executed with the amount of funds allocated for real execution pro-portionally to returns gained in virtual execution on the basis of individual assets and strategies.

# 3 Experimental results

## 3.1 Multi-asset multi-strategy adaptive portfolio management

In order to explore the possibility of using the suggested multi-asset and multi-strategy adaptive active portfolio management agent architecture on the crypto market, we ran backtesting experiments on three months of historical data from the Binance exchange, including September, October, and November of 2021. The data was represented by a full record of historical trades, as well as per-minute LOB snapshots. Four assets, namely BTC, ETH, AAVE, and UNI, were selected for the experiment, with market dynamics presented in Fig. 4.

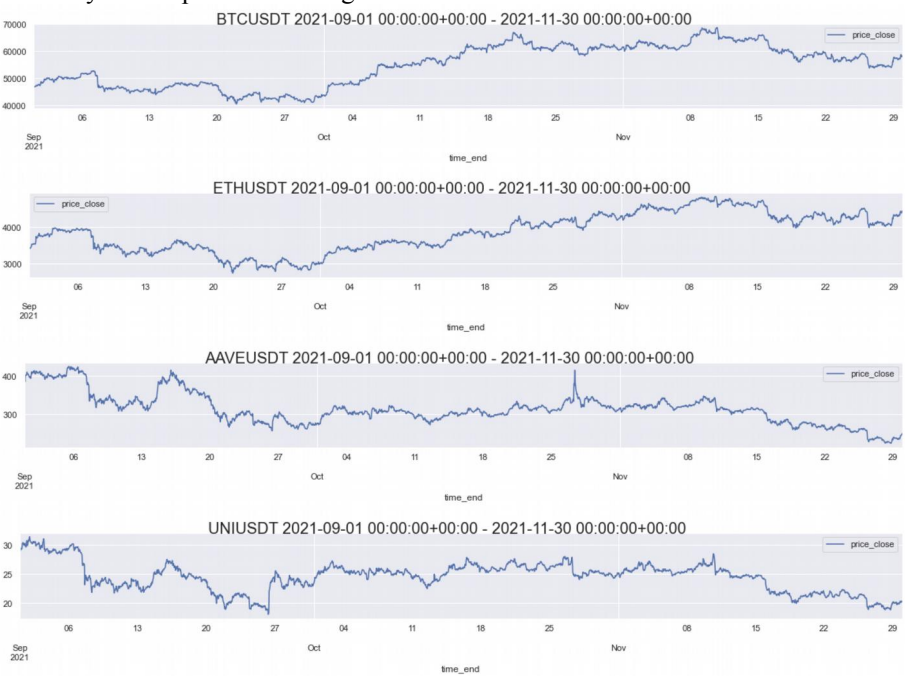

**Fig. 4.** Market dynamics for BTC, ETH, AAVE, and UNI cryptocurrencies during September, October, and November of the year 2021.

The backtesting experiment was performed on the data indicated above with an hourly order refresh rate, with a few different portfolio setups, and cumulative results presented on Fig. 5. One setup was just trying plain single-asset AMSA experiments for each of the four cryptocurrencies individually. Another setup involved a two-asset portfolio of BTC and ETH. The third setup involved a four-asset portfolio, including all four cryptocurrencies. For each of these setups, different time intervals for strategy evaluation and different weighing policies were employed. The intervals for strategy evaluation were 1, 3, 5, 7.5, and 15 days, spanning over respective 90 days of the three months. Two alternative weighing policies were employed. The first policy was evenly splitting the current portfolio fund value across assets and strategies on every iteration of strategy evaluation, for every asset and strategy combination that has rendered a positive return on the previous iteration, denoted as "fixed" on Fig. 5. The second policy was to weight the share of the entire portfolio fund value across asset and strategy combinations proportionally to the value of their positive returns, denoted as "weighted" on Fig. 5.

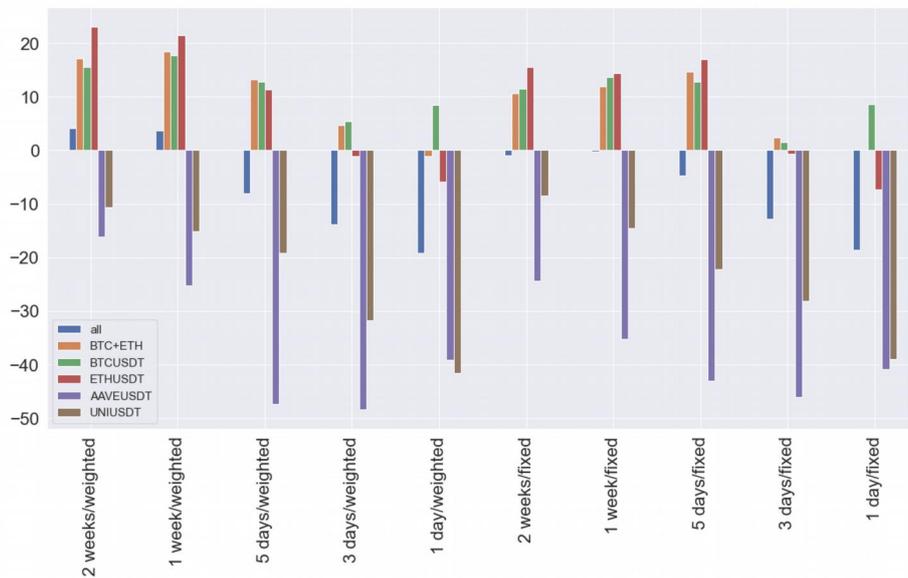

**Fig. 5.** Percentages of returns-on-investment (ROI) for multi-asset adaptive active portfolio management through multi-strategy backtesting on historical data for different types of portfolios rendered as different bars in each bucket (all - portfolio of BTC, ETH, AAVE and UNI; BTC+ETH - portfolio of two assets, other four bars are single-asset). The left five buckets correspond to "weighted" fund allocation on asset/strategy grid, and the right five buckets - for "fixed" allocation. Each five buckets on the left and right correspond to different durations of periods of strategy evaluation and execution iterations.

Interpretation of the results on Fig.5 leads to the following conclusions. First, the "weighted" fund allocation appears more efficient, delivering up to 20% ROI in the case of weekly and bi-weekly strategy evaluation for the two-asset portfolio of BTC and ETH. Second, the weekly and bi-weekly strategy evaluation periods appear superior over the shorter ones. Third, only the combination of "weighted" fund allocation and longer strategy evaluation periods makes it possible to obtain positive returns in the case of a portfolio consisting of all four assets. Fourth, only the combination of the two main high-liquidity coins (BTC+ETH) in the portfolio has provided a non-negative ROI regardless of the other experiment settings, having the performance of the portfolio typically as the average of individual performances of its ingredients, with the exception of the case of the 3-day "weighted" setup where the BTC+ETH portfolio performance has turned out to be superior over the ingredients. At the same time, adding low-liquidity alt-coins to the portfolio was damaging ROI in all cases.

## 3.2 Experiential learning based on simulation and backtesting

The following experiment was run on the same interval of data as described in the previous section, focusing on the BTC/USDT market only. The experiment dealt with per-hour and per-minute market data sampling and order refresh rate during backtesting. Multiple agents employing different strategies were run concurrently in the backtesting environment, relying on the historical data used to simulate real exchange operation, as described in earlier works such as [Raheman, AGI 2021], [Raheman, KNOTH 2021], and [Raheman, 2023]. Each strategy was indicated by order refresh rate (1 hour or 1 minute), bid/ask spread (0.1%, 0.5%, 1%, 2%, 10%), and order cancellation threshold (0%, 0.01%, 0.1%, 1%, 10%). Daily returns (ROI) of each strategy were evaluated, and at the same time, average values of every metric derived from raw market data were computed every day.

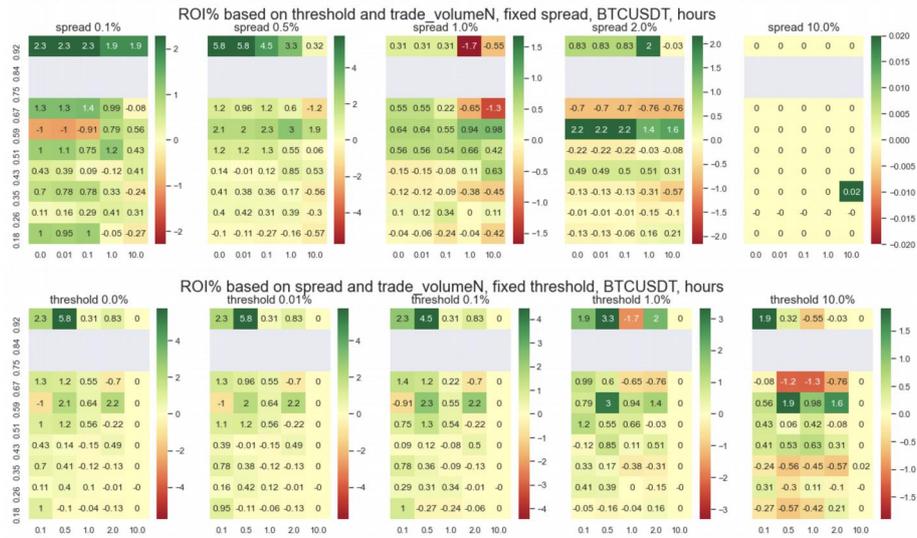

**Fig. 6.** ROI% as a function of strategy parameters (bid/ask spread and order cancellation threshold) and market conditions (normalized trade volume referred to as "volumeN" here) rendered as 2-dimensional slices of a 3-dimensional ("spread" vs. "threshold" vs. "volumeN") cube, displaying the "spots of profit" corresponding to the highest ROI values (such as "spread" at 0.5 for "threshold" up to 1% and "volumeN" above 0.9).

Collecting daily returns per strategy parameters on a daily basis aligned with daily evaluations of the metrics corresponding to specific market conditions made it possible to stack up average ROI numbers in a multi-dimensional space of market strategies and market metrics over 90 days of operations on the exchange. Every point in such space could be further analyzed as a point of either loss or profit, depending on the stacked ROI value at that point. An example of such analysis for a space dimensionality reduced down to a 3-dimensional space is presented in Fig. 6.

The most informative market metrics have appeared to be the standard deviation of the market price, the imbalance between volumes of orders on ask and bid sides of the limit order book (LOB), the imbalance between volumes of trades of buy and sell side, the imbalance between the volume of all trades against the volume of all limit orders, and finally the normalized volume of trades. The latter one is presented as an example on Fig. 6, suggesting that the most profitable spot for market-making is associated with excessively high volumes of trades, spread around 0.5%, and a cancellation threshold up to 1%.

### 3.3 Predictive adaptive market making

The other experiment was run on the latest market data for BTC cryptocurrency during October and November of 2022, as shown in Fig. 7.

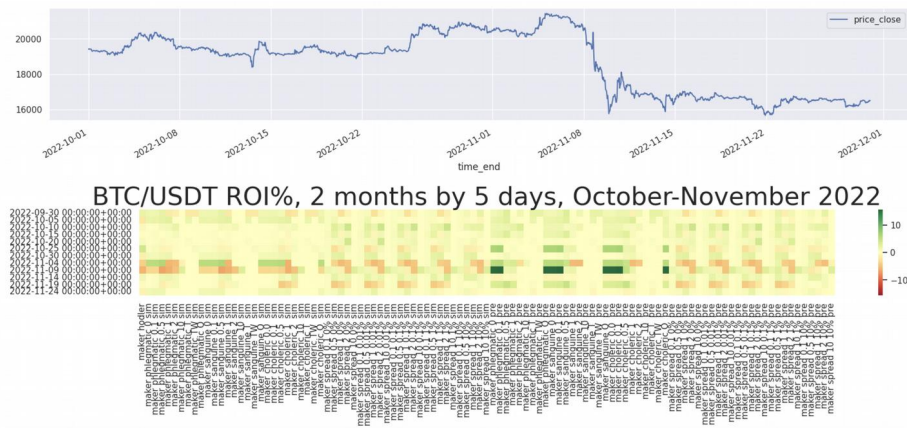

**Fig. 7.** Market dynamics of Bitcoin (BTC) cryptocurrency during October and November 2022 (top) and a heat map of returns and losses per strategy, with a strategy evaluation period of 5 days (bottom). The vertical axis of the heat map corresponds to 12 intervals of 5-day strategy evaluation periods over the 60 days, top to bottom. The horizontal axis of the heat map corresponds to different strategies. The strategies based on experienced price movements are on the left half, while strategies relying on predicted price movements are on the right half. It is clearly seen that in the case of the period associated with a market crash (fourth row from the bottom), non-predictive strategies (left) are losing, while predictive strategies (right) are gaining great profits.

The same family of strategies as in the previous experiment was used, but each strategy was implemented in two different ways by independent agents. The agents of the first kind were handling limit orders based on the current market price and its movements. The agents of the second kind were handling their orders based on anticipated movements of the market price, relying on price predictions projected according to findings presented in earlier works on social media analysis and causal inference [Vishwas, 2022] and [Kolonin, 2023]. The experiment has been run within the same AMSA agent setup and simulation and backtesting framework as described above, with different strategy evaluation periods (15, 10, 5 days), order refresh rate (days, hours), and fund allocation policy ("fixed" and "weighted"), with results presented in Fig. 8.

It has been found that adaptive multi-strategy market-making relying on market price predictions turns out to be rather profitable (up to 25% ROI in 2 months) compared to the same family of strategies being executed without access to predictions, with one exception to one case when fixed fund allocation with 5-day strategy evaluation and daily order refresh rate period has provided 2.5% ROI even without predictions.

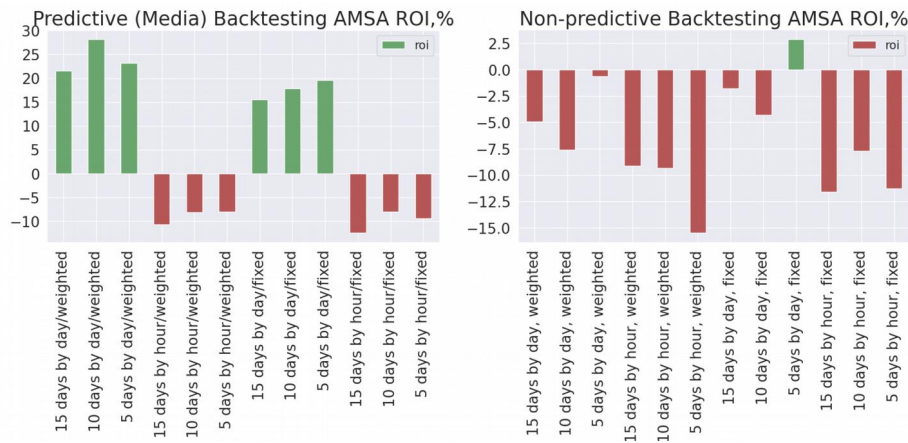

**Fig. 8.** ROI% of adaptive multi-strategy market-making for BTC during October and November 2022, predictive (based on social media) strategies on the left, non-predictive ones on the right.

## 4    Conclusion and future work

Primarily, we have found that the concept of adaptive multi-strategy market-making can be upscaled to active portfolio management for the purpose of risk mitigation. In our future work, we plan to extend it with more strategies involved, including conventional trading based on short and long positions. We also plan to have a more reliable evaluation of the approach or a richer list of assets for longer time periods.

Also, we have explored how to perform experiential learning on the virtual exchange environment simulated by means of backtesting against real historical market data. It has become possible to find meaningful connections between market-making strategies, market conditions, and profits or losses associated with them. Our future work will be dedicated to making this study cover a wider range of assets and financial strategies.

Finally, we have confirmed the value of market price predictions based on social media data on the course of market-making in the simulated environment of backtest-ing. In our future work, we plan to confirm its performance by means of market-making on real-time exchange data.